\documentclass[%
 reprint,
 amsmath,amssymb,
 aps,
floatfix,
]{revtex4-2}

\usepackage{graphicx}% Include figure files
\usepackage{dcolumn}% Align table columns on decimal point
\usepackage{bm}% bold math
\begin{document}

\preprint{APS/123-QED}

%\title{Edge states, transport and topological properties of heterostructures in the SSH model}%

\title{Transversal transport and topological properties of binary heterostructures of topological insulators}%


\author{Rafael Pineda M.}
 %\altaffiliation[raapinedame@unal.edu.co ]{Departamento de física, Universidad Nacional de Colombia}%Lines break automatically or can be forced with \\
\author{William J. Herrera }%
% \email{jherreraw@unal.edu.co}
\affiliation{Departamento de Física \\
Universidad Nacional de Colombia, Bogotá D.C.,111321 Colombia
}%


%\affiliation{
%\affiliation{
 %Third institution, the second for Charlie Author
%\affiliation{%


\date{\today}% It is always \today, today,

\begin{abstract}

 This paper discusses the topological and transport properties of binary heterostructures of different topological materials. The creation of multilayer devices is an alternative to building synthetic topological materials. By adjusting the pattern of layers, we control the global topological properties that favor tunneling and optimize the conductance of the edge state. Using a one-dimensional model and the method of Green's functions, we characterize each layer's edge states and make the chains couplings to generate the heterostructure. To study the bulk properties, we calculate the topological invariant from the Zak phase to build phase diagrams, and we obtain an analytical result for the separation line between different phases that depends on the hopping parameters of the heterostructure. We calculate the differential conductance with the non-equilibrium Green function technique showing the tunneling of the edge states and discussing its possible design and experimental application. 
%\begin{description}
%\end{description}
\end{abstract}

\maketitle

%\tableofcontents

\section{\label{sec:level1}Introduction }

Progress in condensed matter physics opens the possibility of introducing new emergent properties of quantum nature, such as topological insulators (TI) and topological superconductors (TSC), which are classified as a new quantum phase of matter \cite{RevModPhys.83.1057, kane3, PhysRevB.76.045302, RevModPhys.82.3045, KANE20133, asboth2016short, shen2012topological}. The symmetries of the system allow us to define quantities that are invariant associated with the so-called Berry–Pancharatnam–Zak phase, which is a geometric phase of the eigenstates each band \cite{blanco2020, palumbo2019tensor, article4,ghatak2019new}. The metallic states on the surface or edges are protected by time reversal symmetry, making them robust against non-magnetic impurities; therefore, its implementation in quantum computing devices could solve the decoherence problem.  \cite{article3, electronics7100225, PhysRevLett.119.106602}. The helical propagation has spin-momentum locking properties for application in heterostructures with magnetic materials in technologies such as spintronics \cite{PhysRevB.95.205422, electronics7100225, CHONG2021100939, Rachel_2018}.\\

%Similar to the quantum hall effect, topological insulators (TI) present helical electronic states at their surface but without the use of strong external magnetic fields \cite{kane3,RevModPhys.82.3045, asboth2016short}.

A large number of topological materials have been reported in recent years. Quantum Hall effect experiments have derived 2D materials with a high spin-orbit interaction, such as the family of silicenes or germanene, report Dirac zero modes with electronic mobility similar to that of graphene ($10^{6} m/s$)\cite{electronics7100225}. Heterostructures like $HgTe/(Hg, Cd)Te$  and $InAs/GaSb$ quantum wells are also examples of 2D topological materials \cite{kane3, 2016,ren2020engineering, graf2013bulk, gusev2011transport, culcer2010two}. Three-dimensional materials like $Sb_{2}Te_{3}$ or $Bi_{2}Se_{3}$ belong to the family of the strong spin-orbit coupling (SOC) TIs \cite{PhysRevB.95.205422, zhang2010first, manjon2013high}. Crystal lattices are the so-called quintuple layers (QL) which makes them useful for controlling the thickness in the synthesis \cite{Claro2021,ajeel2017topological}. In this frame, the $In_{2}Se_{3}$ is of great interest since it belongs to the same structural family of $Bi_{2}Se_{3}$ of QLs. Still, measurements at the edge by ARPES and low energy models show a trivial insulator \cite{collins2020electronic, li2018large, li2021low}, this makes them relevant materials to build Van der Walls heterostructures with combined topological properties \cite{PhysRevB.95.064302, Rachel_2018, PhysRevA.98.043838, PhysRevB.95.205422, CHONG2021100939, PhysRevLett.108.220401, hoffman1989negative, ballet2014mbe}. In a TI layer, the number of QLs defines the electronic mobility at the edge due to the hybridization of the states of each edge \cite{lang2013competing, wang2011topological}. Magnetoresistance and conductance measurements show a drop in the surface current of thin layers of $Bi_{2}Se_{3}$  with less than 4 QLs, which correspond to an approximate thickness of $4 nm$  \cite{brahlek2015transport, lu2016weak, park2010robustness}. In the case of a few QLs, hybridization between the states increases, and this causes gaped energy bands that behave as a trivial insulators.

\vspace{0.5cm}

Nanotubes, nanowires, or low-dimensional heterostructures are one-dimensional examples of TI and TSC. In the Su-Schrieffer-Heeger (SSH) tight-binding model for insulators, the Peierls distortion generates topologically protected solitonic states \cite{PhysRevB.97.115143, RevModPhys.83.1057, asboth2016short, shen2012topological,li2015winding}. % In superconducting systems, the  topological study predicts isolated pairs of Majorana zero modes, on one-dimensional p-wave superconductor described by Kitaev chain model \cite{RevModPhys.83.1057, asboth2016short, qi2010chiral, xu2014artificial}. These states are of great interest for the development in quantum computing which projects them into future technological applications.  \cite{RevModPhys.83.1057, KANE20133, qi2010chiral, article3, electronics7100225, article4}. 
However, the difficulty in controlling the specific conditions synthesizing of these materials makes Van Der Walls heterostructures an alternative to building them artificially \cite{PhysRevB.95.205422}. For example, Shibayev et al. \cite{Shibayev2019} show experimental topological phase diagrams by measurements of low-temperature magneto-transport for weak antilocalization effects in a binary superlattice of $Bi_{2}Se_{3}$ and $In_{2}Se_{3}$. A similar idea is presented in \cite{Li2014}, where crystalline topological insulators such as $SnTe/CaTe$ build superlattices with multiple valley Dirac states, \cite{li2013single, tanaka2012experimental, hsieh2012topological}. According to the hybridization between the surface states and the superlattice configuration, they suggest the artificial synthesis of weak and strong topological insulators. This motivates us to build a detailed model through the coupling of different chains to study the local effects of the edge states and the interaction between materials with othe topological properties. 
%It would be interesting to do an analytical and numerical analysis of the transport properties through the differential conductance. 
We propose a method to study binary heterostructures of three-dimensional topological insulators based on SSH chains in which we analyze different configurations that combine topological and non-topological materials. In particular, we calculate topological phase diagrams as a function of coupling parameters in SSH chains and binary superlattices with chains of different sizes, as well as the detailed characteristic edge states and differential conductance (DC). For this purpose, we find the Green function (GF) of these heterostructures in a maximally localized Wannier functions basis \cite{RevModPhys.84.1419} to obtain the local density of states (LDOS) that allows characterizing their edge states and study the transport properties.\\

The article is organized as follows. In section 2, we calculate the GF of an arbitrary structure using the SSH model, and the GF and edge states of a heterostructure of chains coupled with different topological properties, in which we study the LDOS. In section 3, we find the topological phase diagrams from the eigenstates of the bulk Hamiltonian, and we show an analytical relationship of the coupling parameters that allows finding the phase diagram. Finally, in section 4, we discuss the transport properties of the edge states using the non-equilibrium Green's function formalism to find the DC, where we analyze the tunneling of the edge states through the heterostructure.

%Starting from the an spinless Hamiltonian in a localized base, the Green function of the primitive cell is calculated and with the Dyson method the function of the edge of a chain of arbitrary size is obtained. The same method allows modeling the edge states structures based on chains with alternating topological and trivial properties. From the reading of the edge states, maps of the intensity of the peak are obtained for the different parametric configurations. Next we consider the topological phase diagrams obtained from the zak phase calculation for a superlattice written in a multiband Hamiltonian [??]. With the zero energy states where band inversion occurs, an analytical parametric relationship is also proposed that allows us to see the phase transition of the proposed structures, where the differential conductance of the structures is calculated, finding a high transmission of the edge states in the tunnel boundary.

\vspace{0.5cm}

\section{\label{sec:level2}Transverse cross section of heterostructure model and Green functions}

The 3D topological materials, such as  $Bi_{2}Se_{3}$ family, have a quintuple-layered structural formation that allows thickness control by using molecular beam epitaxy. The surface has a high electronic mobility given by a metallic state visualized by ARPES. The penetration length of this state in bulk varies according to the composition of the material. In layers with a thickness proportional to the decay length, this effect becomes important since the states of each edge hybridize. The correlation between the edge zero modes generally opens the gap on the surface, and these states disappear. So a minimum size of the topological insulator layer is required for the metallic edge state to be well-defined. The analysis of the characteristics and the penetration length in the bulk of the edge state can be done through a one-dimensional chain model.\\

To study the heterostructures, we use the SSH model, a one-dimensional nearest-neighbor tight-binding system with two sublattice sites represented by atoms A and B \cite{RevModPhys.83.1057, meier2016observation, asboth2016schrieffer, xie2019topological, zhang2021topological}. The Hamiltonian on the basis of spinless localized  orbitals \cite{RevModPhys.84.1419} of a chain of $N$ cells is given by:

\begin{equation}
H=\sum\limits^{N}_{i}(\varepsilon _{A}c_{Ai}^{\dagger}c_{Ai}+\varepsilon_{B}c_{Bi}^{\dagger}c_{Bi}+vc_{Bi}^{\dagger}c_{Ai}+wc_{Bi}^{\dagger}c_{Ai+1}+h.c.),
\end{equation}

\begin{figure}[ht]
\begin{center}
\advance\leftskip-3cm
\advance\rightskip-3cm
\includegraphics[keepaspectratio=true,scale=0.45]{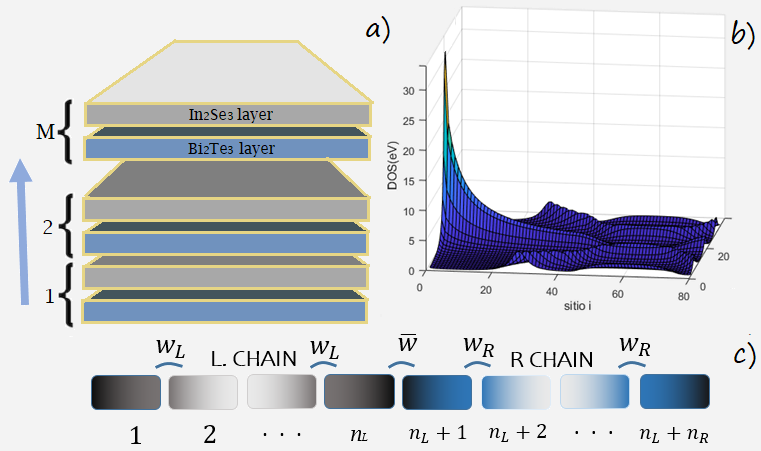}
\caption{a) Schematic representation of a heterostructure of thin layers of alternating topological materials. b) Schematic representation of two chains of size $n_{l}$ and $n_{r}$. The transverse characteristics of the metallic states of the surface are emulated with a one-dimensional model of coupled chains. c) LDOS as a function of cell $i$  of a topological left chain ($w_{L}/v=1.4$) and a trivial right chain ($w_{R}/v=0.8$ and $v_{R}=1$). }
\label{na1}
\end{center}
\end{figure}

\hspace{-0.45cm} where $\varepsilon_{A}$ and $\varepsilon_{B}$ are the self-energies of atoms $A$ and $B$, respectively. The hopping $v$ couples the atoms in the cell, while $w$ couples atoms between neighboring cells. In this system, a topologically protected edge state appears when the intensity of the intercell coupling is greater than the unit cell coupling ($w>v$). The decay length we can control by the values of these parameters and thus emulate the cross-section of the state of a three-dimensional TI. Likewise, it is possible to build an analogous model, as seen in Fig. \ref{na1},  where we show the analogy between a binary superlattice of topological materials and a one-dimensional chain model. 

\bigskip

The cross-section of a heterostructure of $M$ pairs of coupled layers is modeled with a chain whose primitive cell contains the edge state information of the two principal layers. In Fig \ref{na1}  (b), we can see the LDOS of a left topological chain ($w_{L}/v=1.4$) that emulates the behavior of the edge state of a TI layer like $Bi_{2}Se_{3}$ family, with a right trivial chain ($w_{R}/v=0.8$) that emulates the characteristics of a non-topological chain like $In_{2}Se_{3}$, this is schematically represented in Fig \ref{na1} (c). The local GF of the heterostructure is calculated from the coupling of two chains corresponding to the unit cell. This supercell consists of of $N$ molecules composed of two chains $L$ and $R$ of $n_{L}$ and $n_{R}$  cells coupled by hopping $w_{L}$ and $w_{R}$, respectively with $N=n_{L}+n_{R}$. The self-energy $\hat{\Sigma}_{LR}=\hat{\Sigma}^{T}_{RL}$ that couples the two chains at their edges depends on the parameter $\bar{w}$, which we take as the average of the $w_{L}$ and $w_{R}$. By solving Dyson's equation we get the iterative matrix product $\hat{G}=\hat{g}+\hat{g}\hat{\Sigma} \hat{G}$ for each value of energy, we obtain the perturbed GF on the left edge given by:

%In particular, we built a  periodic heterostructure of eight coupled chains ($M=4$).  The system is represented by a supercell composed of two chains, $L$ and $R$, as shown in Fig.\ref{na2} (b). \\

%In order to explore the LDOS of a chain of finite length, we calculate the GF associated to  an isolated molecule with $A$ and $B$ sublattice atoms. 

%In this model, all the  cell couplings are $v$ and all the  intercell couplings $w$ are the same \cite{asboth2016schrieffer}. However, in a more general version, the chain could have different values of hopping of each cell $v_{i}$ and $w_{i}$. Changing the  configurations, we can study more complex models such as superlattice off-diagonal Harper model \cite{lado2019topological}. 

%\begin{equation}
%\hat{g}_{m}(E )=\frac {1} {(E - \varepsilon_A) (E - \varepsilon_B) - v^{2}}\left( 
%\begin{array}{cc}
%E -\varepsilon _{B} & v \\ 
%v & E -\varepsilon _{A}%
%\end{array}%
%\right). 
%\end{equation}

\begin{equation}
\hat{G}_{11}= \hat{g}_{11}+\hat{g}_{1,n_{L}}\hat{\Sigma_{LR}}\hat{G}_{n_{L}+1,1}. 
\label{ge1}
\end{equation}

The non-local GF $\hat{G}_{n_{L}+1,1}$ connects the left edge of the $L$ chain to the left edge of the $R$ chain; this function is calculated using Dyson's equation again:

\begin{equation}
\hat{G}_{n_{L}+1,1}=\hat{g}_{n_{L}+1,n_{L}+1}\hat{\Sigma_{RL}}\hat{G}_{n_{L},1}, 
\label{ge2}
\end{equation}

\hspace{-0.5cm}with

\begin{equation}
\hat{G}_{n_{L},1}= \hat{g}_{n_{L},1}+\hat{g}_{n_{L},n_{L}}\hat{\Sigma_{LR}}\hat{G}_{n_{L},n_{L}+1}. 
\label{ge3}
\end{equation}

Replacing (\ref{ge3}) in (\ref{ge2}) we get the non-local function and replacing in \ref{ge1} we get:

\begin{equation}
\hat{G}_{11}= \hat{g}_{11}+\hat{g}_{1,n_{L}}\hat{\Sigma}_{LR} \hat{\Delta}(\hat{g}_{n_{L},n_{L}},\hat{g}_{n_{L}+1,n_{L}+1}) \hat{\Sigma}_{RL} \hat{g}_{n_{L},1}. 
\label{ge14}
\end{equation}

\hspace{-0.45cm} where

\begin{equation}
\hat{\Delta} (\hat{a},\hat{b}_{})=(\hat{1}-\hat{a}\hat{\Sigma}
\hat{b}\hat{\Sigma}
)^{-1}\hat{a}.
\label{eq5}
\end{equation}

The unperturbed functions $\hat{g}_{11}$ and $\hat{g}_{n_{L},n_{L}}$ in Eq. (\ref{eq5}) are the GF of each edge of the left chain. The operator$\hat{\Delta}(\hat{g}_{n_{L},n_{L}},\hat{g}_{n_{L}+1,n_{L}+1})$ performs the coupling (with self-energy $\bar{w}\hat{C}$) between this chain and the edge of the other chain with the function $\hat{g}_{n_{L}+1,n_{L}+1}$. The non-local GFs  $\hat{g}_{1,n_{L}}$ and $\hat{g}_{n_{L},1}$ appear in Eq. (\ref{ge14})  because we are calculating the GF on one edge, but the coupling with the other chain is done on the other edge. The development of these non-local functions will be necessary for the analysis of transport properties, and they are calculated in appendix A. This allows obtaining the GF at the edge of a heterostructure with a number $M$ of supercells, as shown in Fig. \ref{na1}. From this GF, we obtain the LDOS to analyze the edge states.

\begin{equation}
    \rho_{11}(E)=\frac{-1}{\pi}\operatorname{Im}[Tr(\hat{G}_{11}(E+i\eta ))],
\end{equation}

\hspace{-0.5cm} with $\eta \rightarrow 0$ as  a positive imaginary part in the energy proportional to the energy partition being of the order of $10^{-4}$.\\

%\begin{figure}[ht]
%\begin{center}
%\advance\leftskip-3cm
%\advance\rightskip-3cm
%\includegraphics[keepaspectratio=true,scale=0.7]{ga16.png}
%\caption{Schematic representation of the method that allows us to find the GF at the edge of a chain of arbitrary length. Each molecule is coupled on the left side of the chain with self-energy $\hat{\Sigma}_{12}$. The perturbed GF $\hat{G}_{11}$ depends on $\hat{g}_{11}=\hat{g}_{m}$ and $\hat{g}_{22}$, which is the left edge of the chain.}
%\label{ga16}
%\end{center}\end{figure}

\vspace{0.5cm}

\subsection{Hybridization between the states of each edge}

Now, to analyse the edge states of heterostructure, we calculate the GF at the internal cells of the chain by using Dyson's equation to obtain an LDOS as a function of the cell $j$. To obtain the LDOS in an inner cell of the chain we must build the heterostructure by coupling at site j. A left chain is constructed from $1$ to $j$, and another goes from $j+1$ to $N$. Making the development of the Dyson equation in a similar way to the one we used to generate $G_{11}$ in Eq. (\ref{ge1}) and (\ref{ge2}), we get:

\begin{equation}
 \hat{G}_{j,j}= \hat{g}_{j,j}\hat{\Delta}(\hat{g}_{j,j}\hat{g}_{j+1,j+1})  
 \label{ge4}  
\end{equation}

Where $g_{jj}$ and $g_{j+1,j+1}$ are the GF of the edges of each section that is coupled with selfenergy   $\Sigma_{LR}$, thus, by carrying out this same process for each of the intermediate sites, we can see the edge state's behavior in the chain's internal cell. To study the hybridization between the edge states of each chain, it is helpful to define the non-local GF $\hat{G}_{1N}(E)$, which gives the correlation between the states of the cell $1$ with those in $N$. From this function, we can calculate the probability of propagation of an electron between the edges of the chain and use it in section 4 to calculate transport properties in these systems. We use the same iterative method described in the previous section; a development is made in Appendix A. According to the Dyson equation, the non-local GF  is given by:

%\begin{figure}[ht]
%\begin{center}
%\advance\leftskip-3cm
%\advance\rightskip-3cm
%\includegraphics[keepaspectratio=true,scale=0.55]{gt9.png}
%\caption{(a) LDOS in function of the energy and the cell $n$ of a 30 molecules topological chain with $w/v=1.4$. (b) LDOS at the edge in chains of different sizes where you can see the splitting of the edge state with chains of few cells.}
%\label{gt9}
%\end{center}
%\end{figure}

%On the Fig. \ref{gt9} (a), we observe that the zero energy states of a topological  chain of 30 cells are located at the edges, unlike nonzero energy states that are traveling modes. These states are associated with bulk band systems. The edge  zero energy state loses intensity in the interior according to the length localization $\lambda$, which can be calculated directly from the LDOS  according to the value of the couplings $v$ and $w$ as we see in Fig. \ref{gt9} (c), it is also possible  from a low-energy effective model \cite{xie2019topological}. In this chain, the intensity of the two states is zero after a few molecules in the chain, so they can be considered independent.\\
\bigskip

\begin{equation}
\hat{G}_{1N}(E)=\hat{\Omega} \prod_{i=1}^{N-1}(\hat{\Sigma}\Delta
(\hat{g}_{m}(E),\hat{g}_{ii}(E))), 
\label{eq8}
\end{equation}

\hspace{-0.5cm} with

\begin{equation}
\hat{\Omega}= \hat{\Delta} (\hat{g}_{m}(E),\hat{g}_{m}(E))\hat{\Sigma} 
g_{m}(E).
\end{equation}

The operator $ \hat{\Omega}$  depends only on the GF of one molecule $\hat{g}_{m}(E)$ defined in the equation (2) and with $\hat{\Sigma}_{N-1,N}=\hat{\Sigma}_{12}=\hat{\Sigma}$.  This method has the advantage of preserving the matrix order by increasing chain size, unlike the Hamiltonian approximation, which increases its matrix order according to the length of the chain. This feature is also useful for calculating the Differential Conductance that we will describe in section 4.

\vspace{0.5cm}

\begin{figure}[ht]
\begin{center}
\advance\leftskip-3cm
\advance\rightskip-3cm
\includegraphics[keepaspectratio=true,scale=0.47]{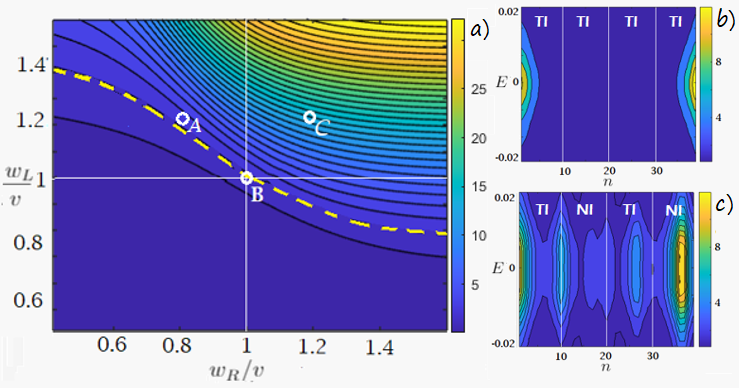}
\caption{a) LDOS at zero energy as function of $w_{L}/v$ and $w_{R}/v$ to define topological and trivial regions similar to phase diagram $n_{R}=3$ and $n_{L}=4$.b) LDOS inside a heterostructure made of two TI supercells ($w_{L}/v=w_{R}/v=1.2$) related to configuration C in (a). c) LDOS inside topological left chain $w_{L}/v=1.2$ and a non-topological right chain ( $w_{R}/v=0.8$) related to configuration A in (a).}
\label{na2}
\end{center}\end{figure}

By analyzing the LDOS at the edge of the heterostructure evaluated at $E = 0$, we can generate intensity maps of this state as a function of the possible configurations of the parameters $w_{L}/v$ and $w_{R}/v$, as observed in Fig.\ref{na2} (b). We have separated the figure's maps into quadrants denoting different configurations. In quadrant IV, both chains are trivial, while in II and III, only one is topological. The LDOS shows a well-defined peak at $E = 0$ in an area delimited by the dashed line in Fig.\ref{na2} (b). In Fig. \ref{na2} (c), (d), we see the LDOS inside a structure of two supercells with configurations of points A and C of the map (b), respectively. When both chains are topological, the internal states disappear due to a weak anti-location effect \cite{li2013single, tanaka2012experimental, hsieh2012topological}. When a topological TI alternates with a trivial chain, the interaction of the edge states strongly depends on the size of each chain and its location length $\lambda$where we can observe the topological zones as will be discussed later. The zones depends on the size of the $n_{L}$ and $n_{R}$ chains.

\section{\label{sec:level3}Topological phase }

In this section, we find the topological invariant of heterostructures from calculating the Berry–Pancharatnam–Zak phase. We impose boundary conditions on the supercell with $c_{N+1}=c_{1}$; this allows us to build phase diagrams by tuning parameters $w_{R/L}/v$ of two chains with sizes $n_{L/R}$ and comparing them with the maps calculated in the previous section. The Zak phase is defined from the occupied states in a discretized First Brillouin Zone as:

\begin{equation}
    \theta=\sum_{i}^{m}\theta _{i}=\sum_{i}^{m}(\log (\prod\limits_{l=1}^{J}\langle \langle
u_{i,k_{l}}|u_{i,k_{l+1}}\rangle\rangle),
\end{equation}

\bigskip

\hspace{-0.43cm} where $m$ is the number of bands $u_{i,k_{l}}$ of the states with energy below the Fermi level $E_{F}$, $k_{l }$ is the momentum variable  in the reciprocal space discretized in $J$ parts. The condition of periodicity imposed on real space allows describing the system in the FBZ with $ k_ {J + 1} = k_ {1} $. The topological invariant takes three characteristic values: zero if the system has a trivial topology and $\pm \pi$  for non-trivial heterostructures. Tuning in the coupling parameters $ w_ {L}/v$ and $w_ {R}/v$ generates a phase diagram that delimits the zones where the non-trivial phase is similar to the maps shown in the previous section.

\vspace{0.5 cm}

\begin{figure}[ht]
\begin{center}
\advance\leftskip-3cm
\advance\rightskip-3cm
\includegraphics[keepaspectratio=true,scale=0.35]{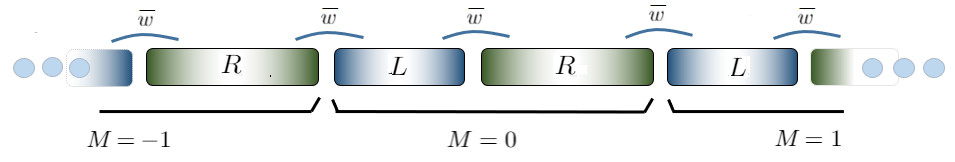}
\caption{Schematic representation of a superlattice made of two chains with different topological properties. }
\label{ga18}
\end{center}\end{figure}

\vspace{0.5cm}

 In Fig.\ref{ga18}, we show a schematic representation of the infinite heterostructure (with an $M>>4$) defined by a primitive cell that is composed of two chains. The change of couplings $ w_ {L}/v$ and $w_ {R}/v$ gives each setting. We perform the analysis in a reciprocal space of a general system of $N$ cells. The Hamiltonian is defined in the following form:  
 
 \vspace{0.5cm}

 $H_{k}=\left( 
\begin{array}{ccccccc}
0 & v_{1} & 0 & 0 & \cdots & 0 & w_{N}e^{-ikL} \\ 
v_{1} & 0 & w_{1} & 0 &  &  & 0 \\ 
0 & w_{1} & 0 & v_{2} & &  & \vdots \\ 
0 & 0 & v_{2} & 0 & \ddots&  &  \\ 
\vdots &  & & \ddots &  \ddots &  & 0 \\ 
0 &  &  &  &  & 0 & v_{N} \\ 
w_{N}e^{ikL} & 0 & & \cdots & 0 & v_{N} & 0%
\end{array}%
\right) $\\

\vspace{1cm}

\hspace{-0.5cm} where $L=Na$ is the size of the supercell. This Hamiltonian has eigenvalues representing the energy bands $ E_ {m} (k) $. We are interested in the band gap close states; in this case, the equation of the system is: 

\begin{equation}
    H |u_{i}\rangle=0.
\end{equation}

From this homogeneous equation, we find a parametric relationship for the cross bands, which, in general, is obtained from the determinant equal to zero det$(H)=0$, and we find that this happens when:

\begin{equation}
   \prod_{i=1}^{N}v_{i}=\prod_{i=1}^{N}w_{i}.
   \label{eq19}
\end{equation}

This result is general for any chain with couplings $v_{i}$ and $w_{i}$ and can be used in our supercell system composed of two coupled chains where $N=n_{L}+n_{R}$. In this case, all the hoppings $v_{i}$  are equal ($v_{i}=v$), and the parameters $w_{i}$ are divided between those of the left and right chain $w_{L/R}$ according to the size of each chain $n_{L/R}$. The parameter $\bar{w}$ that couples the chains is taken as the average between $w_{L}$ and $w_{R}$ $(\bar{w}=\frac{1}{2}(w_{L}+w_{R}))$. Then,  when we have two chains of arbitrary lengths $ n_ {L} $ and $ n_ {R} $, the parametric relationship of Eq. (\ref {eq19})  takes the form:

\begin{equation}
    w^{n_{L}-1}_{L}w^{n_{R}-1}_{R}\bar{w}^{2}=v^{N}.
    \label{eq14}
\end{equation}

\vspace{0.5cm}

%A particular solution of this equation is given by assuming that the parameter that couples both chains is constant $\bar{w}=v$. In this case we have

%\begin{equation}
 %   \frac{w_{L}}{v}=(\frac{v}{w_{R}})^{\frac{n_{R}-1}{n_{L}-1}}. 
%\end{equation}

%\hspace{-0.5cm} This equation shows an inverse relationship in a graph of $w_{L}/v$ vs  $w_{R}/v$. In case the chains are of equal size, the curves are hyperbolas with asymptotes on the axes.\\

%\begin{figure}[ht]
%\begin{center}
%\advance\leftskip-3cm
%\advance\rightskip-3cm
%\includegraphics[keepaspectratio=true,scale=0.75]{gu2.png}
%\caption{Parametric relationship of the hopping $w_{R}$ and $w_{L}$ when hopping between chains is $\bar{w}=v$. The right chain has a fixed size of $n_{R}=8$ coupled with a left chain of different sizes.}
%\label{gu2}
%\end{center}\end{figure}

The algebraic relation of Eq. (\ref{eq14}) is useful as it tells which configurations there is band crossing related to the topological phase transition, allowing us to construct phase diagrams. In the case of the standard SSH model, all couplings $w_{i}$ are the same ($w_{L}=w_{R}=w$), and we obtain the band crossing relation $w = v$ \cite{shen2012topological}. In this way, we find a functional relationship for the coupling parameters of identical systems to those built from the intensity of the edge states at $E=0$. The analytical equation derived from the hamiltonian of the primitive cell in the superlattice generates curves that delimit the configurations with non-trivial topology; this can be seen more quickly if we define the coupling between both chains $\bar{w}=v$  as an approximation:

\begin{equation}
    \frac{w_{L}}{v}=(\frac{v}{w_{R}})^{\frac{n_{R}-1}{n_{L}-1}}
    \label{eq88}
\end{equation}

\begin{figure}[ht]
\begin{center}
\advance\leftskip-3cm
\advance\rightskip-3cm
\includegraphics[keepaspectratio=true,scale=0.67]{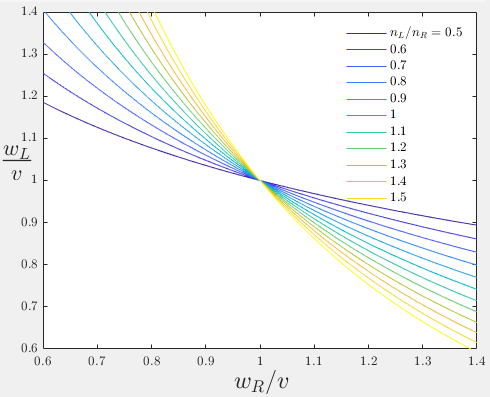}
\caption{Solutions of analytical equation \ref{eq88} as a function of the coupling parameters. Each curve corresponds to the ratio between the size of the left chain and the right chain with $n_{r}$ fixed at 10 cells. These curves represent the parametric phase transition for that chain size ratio.}
\label{gu2}
\end{center}
\end{figure}

This Equation shows the functional relationship between the value of the intercell couplings $w_{L}$ and $w_{R}$. The different curves generated as a function of the ratio between the sizes of each chain ($n_{L}/n_{R}$) are shown in the Fig. \ref{gu2}. Each curve separates the plane into two parts; the upper part corresponds to the topological and the lower to the non-topological. Note that the point of intersection $\frac{w_{R}}{v}=\frac{w_{L}}{v}=1$ delimits the topological phase of each chain, so with $w_{L}<v$ and $w_{R}<v$ both chains are trivial and therefore correspond to a heterostructure configuration with trivial topology.

\begin{figure}[ht]
\begin{center}
\advance\leftskip-3cm
\advance\rightskip-3cm
\includegraphics[keepaspectratio=true,scale=0.37]{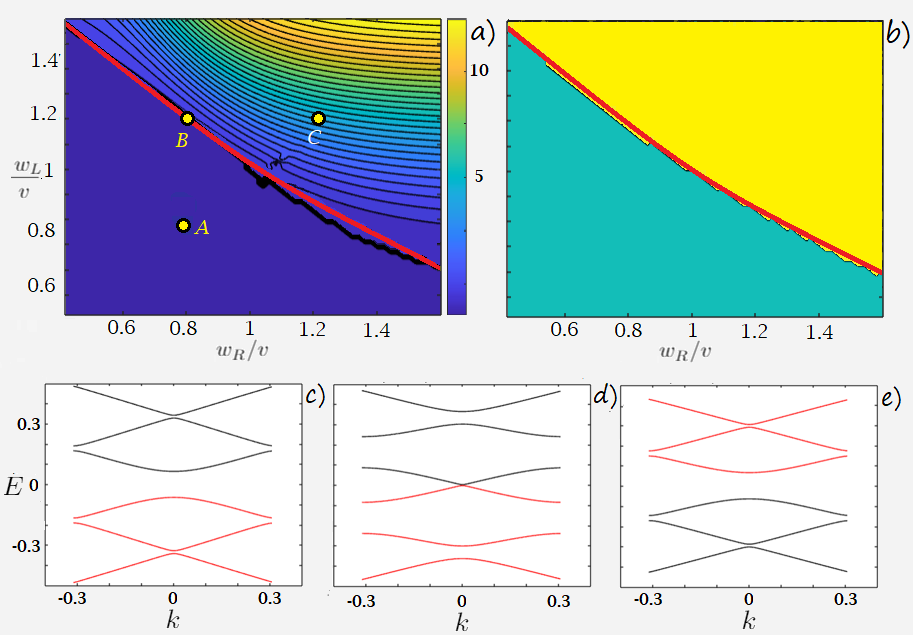}
\caption{(a) Phase diagrams of the edge state intensity in LDOS for $n_{L}=6$ and $n_{R}=4$. (b) Diagram of the topological invariant obtained through the Zak phase. The red line in both maps is obtained from the parametric relationship of band closure of Eq. (\ref{eq14}). (c) (d) and (e) present the electronic spectrum of the superlattice at different points on the diagram. In these spectra, we have bands crossing over the red line that delimit topological region in the phase diagram.}
\label{gt1}
\end{center}\end{figure}

The correspondence of these curves can be verified by overlapping them on the maps obtained in the previous sections. Fig.\ref{gt1} a) presents the intensity map of the zero energy state in the LDOS, and Fig.\ref{gt1} b) of the topological phase diagram obtained from the Zak phase with values of zero in blue, and $\pi$ in yellow. Both maps exhibit similar zones. The red curve on the phase diagram perfectly overlaps the boundary denoted by the diagram. The same curve on the edge state intensity map delimits the zones with a good approximation, as in the Zak phase diagram.

\section{\label{sec:level4}Electric transport }

In this section, we focus on analyzing the local and non-local  differential conductance of heterostructures described in the previous sections; for this, will couple right and left electrodes at the edges of the heterostructure. We model the GF of the edge of each electrode as monatomic semi infinite chains ($v=w\equiv t$) (see appendix A). In non-local transport, there is a potential difference $V$ between the electrodes, generating a current of charge carriers through the heterostructure. In the case of local conductance, we use the left electrode generating a potential difference $V$ with the heterostructure, represented by the GF on edge defined in Eq. (\ref{ge4}). The results derived from this analysis can be contrasted with the phase diagrams of the Fig. \ref{gt1} a) y b). The transport properties are calculated from the formalism of non-equilibrium Green functions and the local and non-local GF already defined in Eq.(\ref{ge4}) and Eq.(\ref{eq8}). The electric current is calculated from the current function, which is given by:

\begin{equation}
   I=\frac{2e}{\hbar}\int dE T(eV)(n_{F}(E-eV)-n_{F}(E)),
\end{equation}

\hspace{-0.5cm} where $n_{F}(E)$ is the Fermi-Dirac distribution, $V$ is the potential difference between the electrodes, and $e$ is the electron's charge. At zero temperature, the differential conductance is proportional to the transmission function given by:

\begin{equation}
    \sigma=\frac{dI}{dV}=\frac{2e}{\hbar}T(eV),
\end{equation}

\hspace{-0.5cm}we obtain the local conductance  $\sigma_{l}$ from the density of states at the edge of a semi-infinite electrode (see appendix A) and the density and GF of the periodic heterostructure.

\begin{equation}
    \sigma_{l}=\frac{8\pi^{2}e}{\hbar}t^{2}_{L}t^{2}_{R}\rho_{L}(eV)\rho_{11}(eV)|\hat{G}_{11}(eV)|^{2}.
\end{equation}

The nonlocal conductance $\sigma_{nl}$ calculates the probability that a charge carrier passes from the left to the right electrode through the structure. This expression includes the LDOS of both semi-infinity electrodes. In this case, the perturbation is made by the couplings with the electrodes, so the nonlocal GF is defined in Eq. \ref{eq8} is an unperturbed function of the system $g_{1N}$. The nonlocal conductance takes the form:

\begin{equation}
    \sigma_{nl}=\frac{8\pi^{2}e}{\hbar}t^{2}_{L}t^{2}_{R}\rho_{L}(eV)\rho_{R}(0)|\hat{G}_{1N}(eV)|^{2},
\end{equation}

\hspace{-0.5cm} With $\rho_{L/R}(E)=-\frac{1}{\pi}$Im(Tr($\hat{G}_{LL/RR}(E+i\eta)$)) corresponds to the LDOS of each electrode, and the non-local GF $\hat{G}_{1N}(eV)$ is perturbed by electrodes with $\Sigma_{L,1}=t_{L}\hat{C}$ and $\Sigma_{N,R}=t_{R}\hat{C}$ (see appendix B). An analytic expression of this function is obtained with the Dyson equation in a similar way to the previous sections and is given by:

 \begin{equation}
 \hat{G}_{1N}
  =\hat{\Lambda}_{L}\hat{g}_{1N}(\hat{1}-\hat{\Lambda}_{R}\hat{\Gamma}_{R}\hat{g}_{N1}\hat{\Lambda}_{L} \hat{g}_{1N})^{-1}\hat{\Lambda}_{R},
  \label{eq21}
 \end{equation}
 
\hspace{-0.5cm} and 

\begin{equation}
\hat{G}_{11} = (\hat{g}_{11}^{-1}-\hat{\Gamma}_{L})^{-1},
\label{eq22}
 \end{equation}
 
\hspace{-0.5cm} with 

\begin{equation}
\hat{\Lambda}_{L}=(\hat{1}-\hat{g}_{11}\hat{\Gamma}_{L})^{-1}\ \ \text{and}\ \hat{\Lambda}_{R}=(\hat{1}-\hat{\Gamma}_{R}\hat{g}_{NN})^{-1}.
\end{equation}

\hspace{-0.5cm} 

\begin{figure}[ht]
\begin{center}
\advance\leftskip-3cm
\advance\rightskip-3cm
\includegraphics[keepaspectratio=true,scale=0.8]{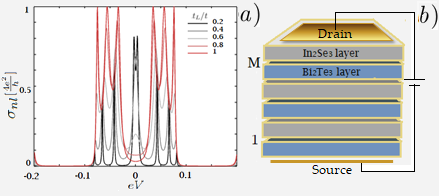}
\caption{a) Non-local conductance of a left topological chain $w_{L}/v=1.2$ and a non-topological right chain ( $w_{R}/v=0.8$) with $n_{L}=6$ and $n_{R}=4$ for different values of the coupling level with electrodes $t_{L}=t_{R}$. b) Schematic representation of the device with the electrodes that allows to measure its non-local conductance}
\label{gh8}
\end{center}\end{figure}

Here $\hat{\Gamma}_{L} = \hat{\Sigma}_{1L} \hat{g}_{LL}\hat{\Sigma} _{L1}$ and $\hat{\Gamma}_{R}=\hat{\Sigma} _{nR}\hat{g}_{RR}\hat{\Sigma}_{Rn}$  are the coupling terms between the structure and the electrodes. The non-local unperturbed GFs $\hat{g}_{1N}$ and $\hat{g}_{11}$ are the ones mentioned in section 2. In general, the non-local GF can correspond to a single chain, a heterostructure, or a binary superlattice. The Fig.\ref{gh8} presents the non-local conductance of a heterostructure of a left topological chain $w_{L}/v=1.2$ and a non-topological right chain ( $w_{R}/v=0.8$) with $n_{L}=6$ and $n_{R}=4$ for different values of the coupling with electrodes $t_{L}=t_{R}$. When the coupling of the structure with the electrodes is weak ($t_{L}/t=0.2$), we are at the tunnel limit  (black curve). This curve describes a conductance with resonances proportional to $N$. The resonances close to $eV = 0$ correspond to a topological edge state given by the parametric relation that defines the topological edge state.

\vspace{0.5cm}

 As shown in Fig.\ref{gh8}, this peak loses intensity when the coupling value between the heterostructure and the electrodes increases. In the transparent limit with a perfect coupling $(t_{L/R}/t=1)$, the intensity of this peak tends to zero (red line), unlike resonances at $eV\not= 0$, which broaden; this indicates that border states have a distinctive characteristics, and their transport occurs only by tunneling. In the tunnel limit, the non-local conductance decays sharply at the transparent limit, which only happens with resonance at $E=0$. As in the LDOS, the DC depends on the size of each chain and the intensity of the couplings that define its topological character.\\

%In heterostructures, the formation of the peak at $E=0$ depends on the intensity of the couplings of each chain and the size of each one, as mentioned in the previous sections. In Fig.\ref{gt6}, we show the electronic transmission curve of two chains with $n_{L}$ and $n_{R}$ sizes, where each curve comes from the change in the parameters $t_{L}=t_{R}$.\\

%\begin{figure*}!
%\begin{center}
%\advance\leftskip-3cm
%\advance\rightskip-3cm
%\includegraphics[keepaspectratio=true,scale=0.6]{gt6.png}
%\caption{Conductance of a heterostructure based on two chains of size $n_{L}=6$, $n_{R}=4$. (a) The largest chain is topological with $w_{L}/v=1.4$, while the shortest one is $w_{R}/v=0.8$. (b)  The largest chain is the non-topological with $w_{L}/v=0.8$, while the shortest one is topological with $w_{R}/v=1.4$. The different curves arise from varying the coupling $t_{L}=t_{R}$ until the transparent limit. }
%\label{gt6}
%\end{center}
%\end{figure*}

\begin{figure}
\begin{center}
\advance\leftskip-3cm
\advance\rightskip-3cm
\includegraphics[keepaspectratio=true,scale=0.41]{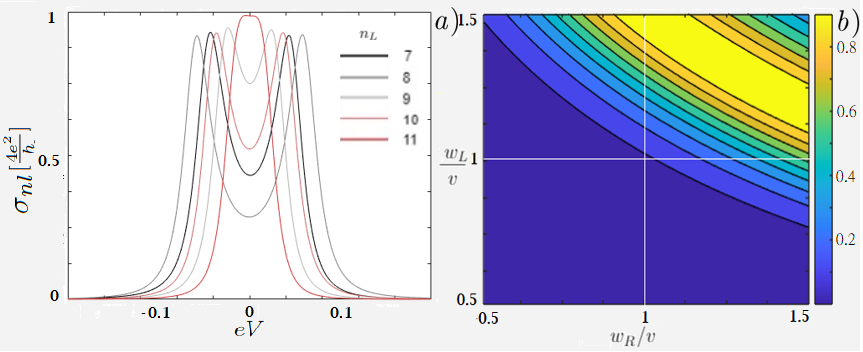}
\caption{(a) Nonlocal conductance in the tunnel limit as a function of voltage. The different curves arise from increasing the length of a left topological chain, with a non-topological chain of a fixed size ($n_{R}=4$) from the right chain. (b) Nonlocal DC intensity map at zero voltage of a binary heterostructure with $M=2$ and ($n_{L}=4$, $n_{R}=6$) tuning the coupling parameters $w_{L}$ and $w_{R}$.}
\label{gt8}
\end{center}
\end{figure}

 Figure \ref{gt8} shows how the conductance of the topological state changes under different parametric conditions. A resonance peak in the conductance of the zero states is formed by increasing the size of the topological chain, which is evidence of the hybridization of the states of each edge as shown in Fig.\ref{gt8}(a). If the chain reaches a sufficient size to define the state, and we continue to increase the size of the chain, the non-local conductance, which is proportional to the function $G_{1N}$, decays with increasing chain length. In Fig.\ref{gt8}(b), we see that there are points of low conductance $\sigma_{nl}(eV=0,w_{L}/v,w_{R}/v)$ in the first quadrant, where both chains of the supercell are topological. As the internal states of each chain interfere with each other, the system is equivalent to that of a effective chain, as we saw in Fig. 4 (c).\\

\begin{figure}[ht]
\begin{center}
\advance\leftskip-3cm
\advance\rightskip-3cm
\includegraphics[keepaspectratio=true,scale=0.4]{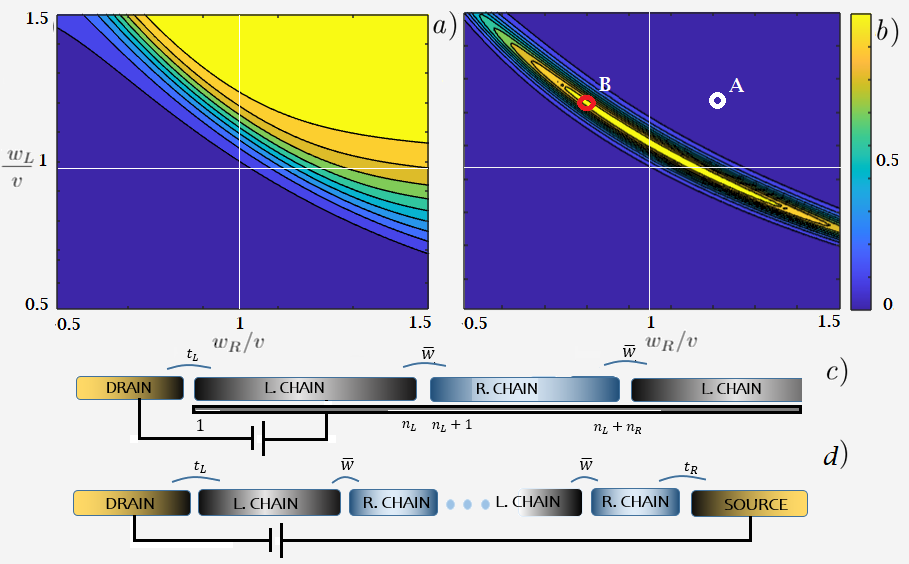}
\caption{(a)Intensity map of the local conductance of the zero energy state,  for different configurations of a heterostructure based on two chains of $n_{L}=6$ and $n_{R}=4$ molecules repeated $M=4$ times alternately. The map is similar to the one obtained for the LDOS in Fig \ref{gt1} . (b) non-local conductance for the same configurations. With the parameters of points A $(0.8,1.2 )$and B $(1.2,1.2)$, the LDOS is calculated close to $E=0$ as shown in Fig. \ref{na2} c) and Fig. \ref{na2} d) respectively. c) schematic representation of the circuit to calculate the local conductance. In this case, only the GF of the edges of the heterostructure is used.}
\label{gu8}
\end{center}
\end{figure}

 In Fig. \ref{gu8} (a) and (b), we can compare local and nonlocal conductance for a binary heterostructure with the same parameters as the one defined in the Fig. \ref{gt8} but repeated four times ($M=4$).  These conductances are represented schematically in Fig. \ref{gu8} (c) and (d) respectively. The local conductance exhibits zones similar to phase diagrams of the invariant in the superlattice; this is due to the periodicity formed by joining the supercell a sufficient number of times, which makes the edge state well defined. This increase in the number of supercells generates a reduction in the areas with nonlocal conductance.  By increasing the number $M$ of primitive cells, these zones are fitted to the parametric line developed with the spectral relation of Eq. (\ref{eq14}); this can be understood from the internal LDOS presented in Fig. \ref{na2} c) and d). Point A in Fig.\ref{gu8} b) indicates that the local DC is zero in a 4-coupled supercell heterostructure of two topological chains. The LDOS in a region close to $E=0$ for each cell $n$ shows that the edge states are formed only at the edges, as shown in Fig. \ref{na2}(c). The DC tends to zero because the edge state decays rapidly inside the chain, unlike the LDOS with the configuration of point B in Fig.\ref{gu8}, which shows intensity peaks inside each supercell in Fig.\ref{na2}(d); this justifies its high value in DC since it creates a collective effect that allows edge states to tunnel through the heterostructure.

\section{Conclusions}

We have studied the topological and transport properties of chain-based binary heterostructures of the SSH model, which emulate the transverse behavior of thin multilayer models based on three-dimensional TI. We have verified the parametric correspondence between an infinite system's topological phase and the edge state's obtained from the LDOS. From the Green function formalism and the Dyson equation, we have calculated the LDOS at the edge of the heterostructure. In different configurations, we have found a peak at zero energy representing a topologically protected metallic state. The intensity of this peak depends on the length of the chains $n_{L}$ and $n_{R}$, which determines hybridization between the states of each edge in a topological chain. For chains with few molecules, the edge state is not defined; this also depends on the relationship between the couplings $w_{R}/v$ and $w_{L}/v$ in the supercell.  With the intensity peak at zero-energy state tuning the coupling between molecules, we have built maps of the different parametric configurations. The LDOS map shows a curve that separates topological heterostructures from those that are not.\\

The topological invariant obtained from the Berry–Pancharatnam–Zak phase allows parametric configurations with a non-trivial topological phase identification. These zones are similar to those found in the study of the edge state in the LDOS at $E=0$ according to the bulk-surface correspondence. The curve that separates the topological zones from the trivial ones and that marks the parametric phase transition is derived from the
 analytical relationship generated by the crossing of bands in an infinite system. This relationship is useful because we can quickly find the phase diagrams for superlattices with chains of different sizes.

In the transport properties calculated from the non-equilibrium Green functions formalism, we have shown that the non-local differential conductance of these states through the heterostructure occurs only by tunneling, which is a property of topological states. The conductance map at $eV=0$ describes high conductance zones similar to those calculated with the edge states and the topological phase. However, in heterostructures of several coupled supercells, the configurations that give high conductance are reduced due to the increased dimensions of the system. For the case of heterostructures with several supercells, the DC at $E=0$ can be used to find the topological phase transition since, near the transition line, the edge states decay slowly, and the conductance shows a peak at $eV=0$.  We expect that our analysis can guide future experiments for studing heterostructures based on alternating different topological materials.
\\

\textbf{ACKNOWLEDGEMENT}\\

We thank Dr. Pablo Burset for his comments and contributions that significantly improved the manuscript. DIEB, Hermes code 48148 support this work.

\vspace{1.5cm}

\textbf{Appendix A: Green function construction for finite and semi-infinite chain}

\vspace{0.5cm}

In this appendix, we show how to calculate the GF of a finite chain with the Dyson equation. We also derive an analytic expression for the GF of a semi-infinite chain. The process starts with molecule 1 GF $\hat{g}_{m}(E )$ being perturbed by the edge of a molecule or a chain of some size with unperturbed GF $g_{22}$ and coupling $\Sigma_{12}$:

\begin{equation}
\hat{g}_{m}(E )=\frac {1} {(E - \varepsilon_A) (E - \varepsilon_B) - v^{2}}\left( 
\begin{array}{cc}
E -\varepsilon _{B} & v \\ 
v & E -\varepsilon _{A}%
\end{array}%
\right). 
\end{equation}

\bigskip

For simplicity, we take self-energy values equal to zero ($\varepsilon_{A} = \varepsilon_{B}=0$). To couple two of these molecules denoted by $\hat{g}_{11}$ and $\hat{g}_{22}$,  we use the Dyson's equation approach through self-energy of the form: 

\begin{equation}
\hat{\Sigma}_{12}=w\left( 
\begin{array}{cc}
0 & 0 \\ 
1 & 0 %
\end{array}%
\right)=w\hat{C} ,
\end{equation}

\bigskip

\hspace{-0.5cm} where $\hat{\Sigma}_{12}=\hat{\Sigma}^{T}_{21}$. By solving the Dyson equation, which numerically reduces to the iterative matrix product $\hat{G}=\hat{g}+\hat{g}\hat{\Sigma} \hat{G}$ for each value of energy , we obtain the perturbed GF on the left edge given by

\vspace{0.5cm}

\begin{equation} hat{G}_{11}=\hat{g}_{11}+\hat{g}_{11}\hat{\Sigma} _{12}\hat{G}_{21} 
\end{equation}

\hspace{-0.5cm} The perturbed GF $\hat{G}_{11}$ depends on the non-local GF $\hat{G}_{21}$ that should be calculated similarly. By applying the Dyson equation again and knowing that unperturbed GF $\hat{g}_{21}$ is zero since before the coupling, both molecules were separated, the following expression is achieved\\

\vspace{0.5cm}
$\hat{G}_{21}=\hat{g}_{22}\hat{\Sigma} _{21}\hat{G}_{11},$

\vspace{1cm}

\hspace{-0.5cm} then replacing:

\begin{equation}
    \hat{G}_{11}(E)=(\hat{1}-\hat{g}_{11}\hat{\Sigma} _{12}\hat{g}_{22}\hat{\Sigma} _{21})^{-1}\hat{g}_{11}=\hat{\Delta} (\hat{g}_{m},\hat{g}_{22}).
    \label{ap1}
\end{equation}

$\hat{\Delta}$ is an operator that depends on the GF of each structure to be coupled. In this case, we find the perturbed GF of two coupled molecules, so $\hat{g}_{11}=\hat{g}_{22}\equiv\hat{g}_{m}$. When coupling a new molecule to the chain, the unperturbed GF replaces it with the one calculated in the previous perturbation $\hat{g}_{22}=\hat{G}_{11}$, while $\hat{g}_{11}$ is always the GF of one molecule $\hat{g}_{m}$; this is how we generate a system of arbitrary size through the GF on edge:\\

%The schematic representation in Fig.\ref{ga16} shows the method  for coupling another cell to the chain. We initially calculate the perturbed GF $\hat{G}_{11}$ of coupling a molecule denoted by $\hat{g}_{11}$ to the left side of a chain with $\hat{g}_{22}$. 

\begin{figure}[ht]
\begin{center}
\advance\leftskip-3cm
\advance\rightskip-3cm
\includegraphics[keepaspectratio=true,scale=0.7]{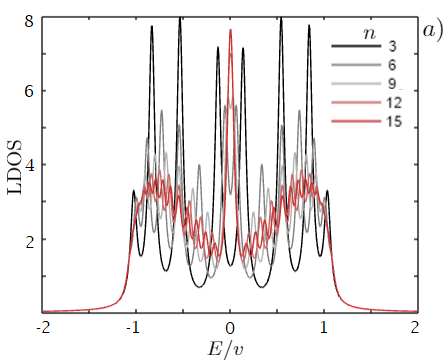}
\caption{(a) LDOS at the left edge of a finite topological chain with different sizes ($w/v=1.2$).}
\label{gt11}
\end{center}
\end{figure}

\vspace{1cm}

\vspace{0.5cm}

\hspace{-0.5cm} The Eq. \ref{ap1} can be solved analytically in a semi-infinite chain where $\hat{G}_{11}=\hat{g}_{22}$. In this case, we obtain a Green function of the form:

\vspace{0.5cm}

\begin{equation}
G_{11}(E)=\frac{D}{w}\left( 
\begin{array}{cc}
\frac{w}{AD}\left[ 1+\frac{v}{w}D\right] & 1 \\ 
1 & \frac{A}{v}%
\end{array}%
\right),
\end{equation}

\vspace{0.5cm}

\hspace{-0.5cm} with $A/B=(E - \varepsilon_{A/B})$. $D$ is an armonical function of energy and the parameters $v$ and $w$ 

\begin{equation}
 D=\frac{AB-(v^{2}+w^{2})}{2vw}-i\sqrt{1-\left[ \frac{%
AB-(v^{2}+w^{2})}{2vw}\right] ^{2}}.  
\end{equation}

\vspace{0.5cm}

The LDOS of the semi-infinite chain can be observed in Fig. \ref{gt11} the red line. To find the non-local GF, we must do a recursive calculation. The perturbed GF $\hat{G}_{1N}$ arises from coupling cell $N$ with a chain of $N-1$ molecules. In a small system for $N=3$, for example, we have:

\vspace{1cm}

\begin{center}
    $\hat{G}_{13}=\hat{g}_{12}\hat{\Sigma} _{23}\hat{G}_{33},$

\bigskip

$\hat{G}_{33}=\hat{g}_{33}+\hat{g}_{33}\hat{\Sigma} _{32}\hat{G}_{23},$

\bigskip

$\hat{G}_{23}=\hat{g}_{22}\hat{\Sigma} _{23}\hat{G}_{33}.$
\end{center}

\vspace{1cm}

\hspace{-0.5cm} From here, we have:

\vspace{0.5cm}

\begin{equation}
  \hat{G}_{33}=(\hat{1}-\hat{g}_{33}\hat{\Sigma}_{32}\hat{g}_{22}\hat{\Sigma}_{23})^{-1}\hat{g}_{33},  
\end{equation}

\bigskip

\hspace{-0.5cm} By substituting, we get the non-local GF:

\begin{equation}
\hat{G}_{13}=\hat{g}_{12}\hat{\Sigma}_{23}(1-\hat{g}_{33}\hat{\Sigma} _{32}\hat{g}_{22}\hat{\Sigma}_{23})^{-1}\hat{g}_{33}, 
\end{equation}

\vspace{0.5cm}

\hspace{-0.5cm} We still need to calculate functions $\hat{g}_{12}$ and $\hat{g}_{22}$, with a similar functional form:

\begin{equation}
    \hat{G}_{12}=\hat{g}_{11}\hat{\Sigma} _{12}(\hat{1}-\hat{g}_{22}\hat{\Sigma} _{21}\hat{g}_{11}\hat{\Sigma}
_{12})^{-1}\hat{g}_{22}.
\end{equation}

\bigskip

\hspace{-0.5cm} More generally, for a chain of $N$ molecules, we can write the GF as:

\begin{equation}
\hat{G}_{1N}=\hat{g}_{1,N-1}\hat{\Sigma}(\hat{1}-\hat{g}_{NN}\hat{\Sigma}\hat{g}_{N-1,N-1}\hat{\Sigma})^{-1}\hat{g}_{NN},
\end{equation}

with

\begin{equation}
 \hat{G}_{1N}=\hat{g}_{1,N-1}\Delta(\hat{G}_{N-1,N-1}),   
\end{equation}

\hspace{-0.5cm} and 

\begin{equation}
  \Delta(\hat{g}_{N-1,N-1})=\hat{\Sigma}(\hat{1}-\hat{g}_{NN}\hat{\Sigma}\hat{g}_{N-1,N-1}\hat{\Sigma})^{-1}\hat{g}_{NN}.  
\end{equation}

\vspace{0.5cm}

In this way each GF is computed until performing the calculation of a chain with two molecules:

\begin{equation}
\hat{g}_{1N}=\hat{\Omega}
(\hat{g}_{m})\prod_{i=1}^{N-1}\Lambda
(\hat{g}_{ii}^{r}),
\label{eq32}
\end{equation}

\hspace{-0.5cm} with

\begin{equation}
\hat{\Omega}= \hat{\Delta} (\hat{g}_{m}(E),\hat{g}_{m}(E))\hat{\Sigma} 
g_{m}(E).
\label{eq33}
\end{equation}

\vspace{0.5cm}

\textbf{Appendix B: Non local Green function and conductance relations }

\vspace{1cm}

Here, we show how to calculate the GF of a structure perturbed by edge electrodes, which allows expressing of the differential conductance from a transmission function. We perform the coupling of the non-local GF to the electrodes. We start with the left electrode by applying the Dyson equation to achieve coupling with self-energy $\hat{\Sigma} _{1L}$: 

\vspace{0.5cm}

\begin{equation}
  \hat{G}_{1N}=\hat{g}_{1N}+\hat{g}_{11}\hat{\Sigma} _{1L}\hat{g}_{LL}\hat{\Sigma} _{L1}\hat{G}_{1N},  
\end{equation}

\vspace{0.5cm}

\hspace{-0.5cm} where $\hat{g}_{LL}$ is the GF of a semi-infinite monatomic chain described in appendix A. Defining $A=\hat{\Sigma} _{1L}\hat{g}_{LL}\hat{\Sigma} _{L1}$,

\vspace{0.5cm}

\begin{equation}
 \hat{G}_{1N}=(\hat{1}-\hat{g}_{11}A)^{-1}\hat{g}_{1N}.   
\end{equation}\\

Now, for coupling with the right electrode we proceed in a similar way: 

\begin{equation}
   \hat{G}_{1N}=\hat{g}_{1N}+\hat{G}_{1N}B\hat{g}_{NN},
\end{equation}

\vspace{0.5cm}

\hspace{-0.5cm} By defining $B=\hat{\Sigma} _{NR}\hat{g}_{RR}\hat{\Sigma} _{RN}$, and $\hat{g}_{RR}$, the GF of the right electrode,

\begin{equation}
    \bigskip \hat{G}_{1N}=\hat{g}_{1N}(1-B\hat{g}_{NN})^{-1}.
\end{equation}

\hspace{-0.5 cm} It is worth mentioning that in Eq. (\ref{ge3}), the unperturbed GF $\hat{g}_{1N}$ corresponds to the perturbed function $G_{1N}$ of Eq. (\ref{eq33}) that already includes the left electrode. Finally, we calculate $\hat{g}_{NN}$ , which is equivalent to $\hat{G}_{NN}$ that couples the chain with the left electrode. The calculated GF now becomes unperturbed functions to perform the Dyson of the total coupling:

\vspace{0.5cm}

\begin{equation}
    \hat{G}_{NN}=\hat{g}_{NN}^{c}+\hat{g}_{N1}^{c}A\hat{g}_{1N}^{c}(\hat{1}-A\hat{g}_{11}^{c})^{-1}.
\end{equation}

\vspace{0.5cm}

\hspace{-0.5cm} Here, $g^{c}_{ij}$ are the functions of the structure when it has not been attached to any electrode calculated in appendix A. In summary,

\begin{eqnarray*}
\hat{G}_{1N} &=&\hat{g}_{1N}(1-B\hat{g}_{NN})^{-1} \\
\hat{g}_{1N} &=&(\hat{1}-\hat{g}_{11}^{c}A)^{-1}\hat{g}_{1N}^{c} \\
\hat{g}_{NN} &=&\hat{g}_{NN}^{c}+\hat{g}_{N1}^{c}A\hat{g}_{1N}^{c}(\hat{1}-A\hat{g}_{11}^{c})^{-1}.
\end{eqnarray*}

\vspace{0.5cm}

\hspace{-0.5cm} By Substituting the equations and applying matrix commuting properties, we finally obtain the following:

\begin{equation}
    \hat{G}_{1N}=\left[ ((1-B\hat{g}_{NN}^{c})\left( \hat{g}_{1N}^{c}\right)
^{-1}(\hat{1}-\hat{g}_{11}^{c}A)-B\hat{g}_{N1}^{c}A\right] ^{-1}.
\end{equation}

With this function we calculate the DC described in Eq.(18). This conductance can be expressed in terms of the electronic transmission as shown in \cite{gomez, vasilesca, Paulsson2003ResistanceOA}.

\begin{equation}
    T(eV)=4\pi^{2}t^{2}_{L}t^{2}_{R}\rho_{L}(eV)\rho_{R}(0)|\hat{G}_{1N}(eV)|^{2}.
\end{equation}

%\begin{figure}[ht]
%\begin{center}
%\advance\leftskip-3cm
%\advance\rightskip-3cm
%\includegraphics[keepaspectratio=true,scale=0.4]{ga13.png}
%\caption{a) Transmission at the transparent limit of a topological chain of different leng.nl=11 wl=0.7 wr= 0.4 b) DOS for $ E = 0 $ in function of the internal site of the chain of 20 sites, }
%\label{visina8}
%\end{center}\end{figure}

%% Fundamentos de la fase topológica

\nocite{*}
\bibliography{aipsamp.bib}% Produces the bibliography via BibTeX.

%%%%%%%%%%%%%

\end{document}